\documentclass[RNAAS]{aastex631}

\newcommand{\bprp}{\mathrm{G_{BP}\!-\!G_{RP}}}
\newcommand\meanRhk{\left<R'_{\rm HK}\right>}

\begin{document}

\title{\large Extending the Asteroseismic Calibration of the Stellar Rossby Number}

\author[0000-0003-4034-0416]{Travis S.~Metcalfe}
\affiliation{Center for Solar-Stellar Connections, White Dwarf Research Corporation, 9020 Brumm Trail, Golden, CO 80403, USA}

\author[0000-0001-8835-2075]{Enrico Corsaro}
\affiliation{INAF - Osservatorio Astrofisico di Catania, Via S. Sofia, 78, 95123, Catania, Italy}

\author[0000-0003-3175-9776]{Alfio Bonanno}
\affiliation{INAF - Osservatorio Astrofisico di Catania, Via S. Sofia, 78, 95123, Catania, Italy}

\author[0000-0003-1853-6631]{Orlagh L.~Creevey}
\affiliation{Universit\'e C\^ote d'Azur, Observatoire de la C\^ote d'Azur, CNRS, Laboratoire Lagrange, France}

\author[0000-0002-4284-8638]{Jennifer L.~van~Saders}
\affiliation{Institute for Astronomy, University of Hawai`i, 2680 Woodlawn Drive, Honolulu, HI 96822, USA}

\begin{abstract} % 146 words / 150 limit

The stellar Rossby number (Ro) is a dimensionless quantity that is used in the 
description of fluid flows. It characterizes the relative importance of Coriolis forces 
on convective motions, which is central to understanding magnetic stellar evolution. Here 
we present an expanded sample of Kepler asteroseismic targets to help calibrate the 
relation between Ro and Gaia color, and we extend the relation to redder colors using 
observations of the mean activity levels and rotation periods for a sample of brighter 
stars from the Mount Wilson survey. Our quadratic fit to the combined sample is nearly 
linear between $0.55 < \bprp < 1.2$, and can be used to estimate Ro for stars with 
spectral types between F5 and K3. The strong deviation from linearity in the original 
calibration may reflect an observational bias against the detection of solar-like 
oscillations at higher activity levels for the coolest stars.

\end{abstract}

%%%%%%%%%%%%%%%%%%%%%%%%%%%%%%%%%%%%%%%%%%%%%%%%%%%%%%%%%%%%%%%%%%%%%%%%%% 
\section{1.\ Introduction}

The stellar Rossby number (Ro) is a key parameter that is related to both the magnetic 
evolution of stars and the space weather environment of their planetary systems. From an 
observational perspective, it is traditionally estimated as the ratio of the stellar 
rotation period to the convective turnover time (Ro $\equiv P_\mathrm{rot}/\tau_c$). 
While the rotation period can often be inferred directly from observations, the 
convective turnover time depends on properties of the stellar interior that can only be 
determined indirectly from stellar models. The most widely used approach to estimate the 
convective turnover time was developed more than 40 years ago \citep{Noyes1984}. 
Exploiting the fact that stellar activity levels are correlated with Ro, this approach 
used the chromospheric activity levels and rotation periods for 41 main-sequence FGK 
stars observed by the Mount Wilson Observatory (MWO) survey to fit a semi-empirical 
relation between $\log\tau_c$ and B$-$V color. However, this relation depends on 
relatively indirect constraints from stellar evolution models, and yields values of 
$\tau_c$ for the Sun that are inconsistent with standard solar models.

Recently, a new approach was pioneered using asteroseismic models for 62 stars with 
solar-like oscillations detected by the Kepler mission to calibrate a relation between 
$\tau_c$ and Gaia color \citep{Corsaro2021}. This analysis relied on asteroseismology to 
yield precise values for all of the stellar properties that are required to estimate 
$\tau_c$ directly, including the depth of the surface convection zone as well as the 
stellar radius, mass, and luminosity: \mbox{$\tau_c \simeq d_{cz}(M/LR)^{1/3}$}. The 
approach yields a very good estimate of the local convective turnover time at the base of 
the stellar convection zone, matching the value for the Sun obtained from a fully 
calibrated standard solar model \citep{Bonanno2002}. The primary limitation of this 
relation is the range of Gaia colors for the Kepler asteroseismic sample.

\section{2.\ Expanding the Asteroseismic Sample}

\begin{figure*}[t]
 \centering\includegraphics[width=4.1in]{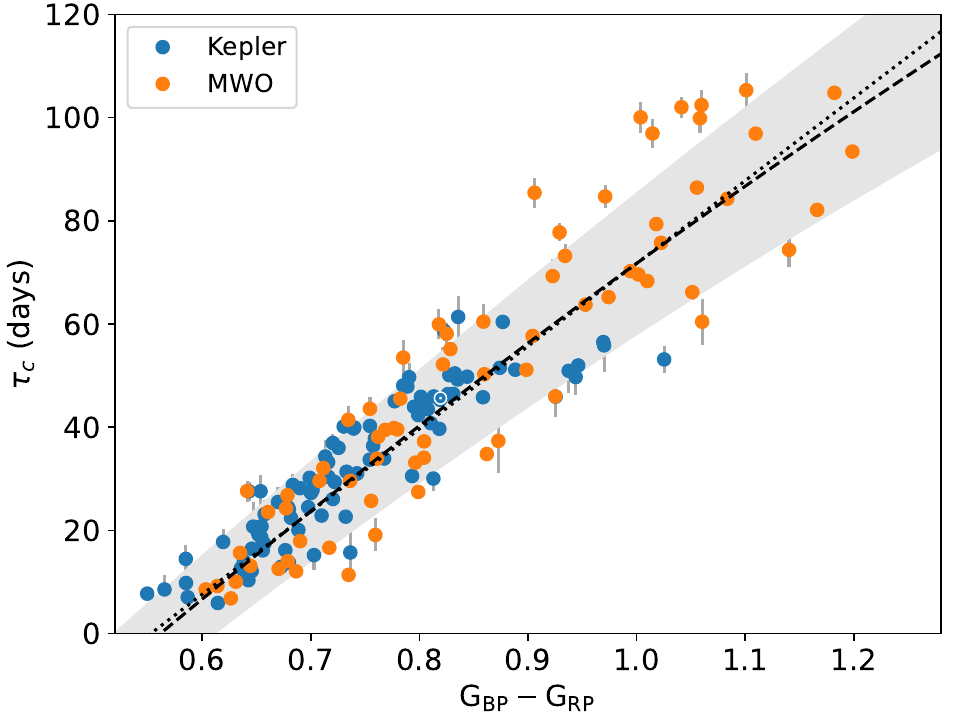}
 \caption{Convective turnover time $\tau_c$ as a function of Gaia color for the expanded 
sample of Kepler asteroseismic targets (blue points) following \cite{Corsaro2021}, and 
inferred for the MWO sample (orange points) from the measured mean activity levels and 
rotation periods published in \cite{Baliunas1996}. The dashed line shows the quadratic 
fit to the combined sample with the 1$\sigma$ credible interval shaded, while the dotted 
line shows the linear fit for comparison.\label{fig1}}
 \end{figure*}

The Kepler mission produced two well characterized samples of asteroseismic targets: the 
LEGACY sample of solar-type main-sequence stars that were observed in short-cadence for 
at least 12 months \citep{Lund2017}, and the KOI sample of suspected and confirmed 
exoplanet host stars with detected solar-like oscillations \citep{Davies2016}. The 
original calibration of $\tau_c$ as a function of Gaia color only used the LEGACY sample, 
because the published modeling results for the KOI sample \citep{SilvaAguirre2015} did 
not tabulate the depth of the surface convection zone. We compiled the necessary details 
from a uniform set of modeling results for the LEGACY and KOI samples, which were 
obtained with v1.3 of the Asteroseismic Modeling Portal \citep[AMP;][]{Metcalfe2009, 
Creevey2017}. This expanded sample increases the total number of stars from 62 to 96, and 
extends the reddest color from $\bprp=0.97$ to 1.02. The resulting values of $\tau_c$ are 
shown as blue points in Figure~\ref{fig1}.

% FIGURE 1 ---------------------------------------------------------------   
%-------------------------------------------------------------------------               

\section{3.\ Extending the Calibration}

For a sample of stars from the MWO survey with well determined mean activity levels 
$\meanRhk$ and rotation periods, \cite{Brandenburg1998} identified a correlation between 
$\meanRhk$ and $\mathrm{Ro}^{-1}$. Inspired by this correlation, we assume that $\meanRhk 
\propto \mathrm{Ro}^{-1}$ and predict $\mathrm{Ro}/\mathrm{Ro}_\odot$ from the inverse of 
the mean activity level relative to solar: $\mathrm{Ro}/\mathrm{Ro}_\odot \simeq 
\meanRhk_\odot/\meanRhk$. We then use $P_\mathrm{rot}$ to infer $\tau_c$ from 
$\mathrm{Ro}/\mathrm{Ro}_\odot$:
 \begin{equation}
 \tau_c \equiv \frac{P_\mathrm{rot}}{\mathrm{Ro}}
        = \frac{P_\mathrm{rot}}{0.496\,\mathrm{Ro}/\mathrm{Ro_\odot}}
        \simeq \frac{P_\mathrm{rot}}{0.496\,\meanRhk_\odot/\meanRhk}. \label{eq1}
 \end{equation} 
Here we have chosen $\mathrm{Ro}_\odot=0.496$ to anchor our calibration of $\tau_c$ to 
the asteroseismic scale of \cite{Corsaro2021}. Using the measured values of $\meanRhk$ 
and $P_\mathrm{rot}$ from \cite{Baliunas1996} for the MWO sample, and adopting 
$\meanRhk_\odot$ from \cite{Egeland2017}, the results of Eq.(\ref{eq1}) are shown as 
orange points in Figure~\ref{fig1}.

Following the same procedures described in \cite{Corsaro2021}, we performed linear and 
quadratic fits to both the combined sample and the MWO sample by itself. The Bayesian 
evidence strongly supports the quadratic model in both cases, indicating a small 
departure from linearity across the observed color range. The fit relation (dashed line 
in Figure~\ref{fig1}) has the same form as Eq.(11) in \cite{Corsaro2021}, $\tau_c = b'_1 
+ b'_2(\bprp) + b'_3(\bprp)^2$~days, with the following parameters: $b'_1 = 
-106.6^{+4.6}_{-4.5}$~d, $b'_2 = 204.4^{+10.9}_{-11.3}$~d~mag$^{-1}$, and $b'_3 = 
-26.1^{+6.6}_{-6.4}$~d~mag$^{-2}$. The shaded region in Figure~\ref{fig1} indicates the 
1$\sigma$ credible interval, and the linear fit is shown as a dotted line for comparison.

\section{4.\ Discussion}

There is remarkable agreement between the asteroseismically calibrated values of $\tau_c$ 
and those inferred from the measured mean activity levels and rotation periods (only the 
same zero-point is enforced by our choice of $\mathrm{Ro}_\odot$). This validates our 
assumption that $\meanRhk \propto \mathrm{Ro}^{-1}$. Our quadratic fit to the combined 
sample is nearly linear between $0.55 < \bprp < 1.2$, and can be used to estimate Ro for 
stars with spectral types between F5 and K3. Notably, the asteroseismic sample appears to 
be confined to the lowest values of $\tau_c$ at the reddest colors. These targets 
generally have higher Rossby numbers and lower activity levels, where the weaker 
intrinsic oscillation amplitudes are least suppressed by magnetic activity. Consequently, 
the strong deviation from linearity in the original asteroseismic calibration may reflect 
an observational bias against the detection of solar-like oscillations at higher activity 
levels for the coolest stars.

%%%%%%%%%%%%%%%%%%%%%%%%%%%%%%%%%%%%%%%%%%%%%%%%%%%%%%%%%%%%%%%%%%%%%%%%%% 
\begin{acknowledgments}
T.S.M.\ acknowledges support from NASA grant 80NSSC22K0475, NSF grant AST-2205919, and 
XSEDE allocation TG-AST090107. A.B.\ and E.C.\ acknowledge support from the INAF grant 
``Unveiling the magnetic side of the stars'' and MIUR grant CHRONOS. ESA Gaia data are 
processed by the Gaia Data Processing and Analysis Consortium. J.v.S.\ acknowledges 
support from NSF grant AST-2205888.
\end{acknowledgments}

%%%%%%%%%%%%%%%%%%%%%%%%%%%%%%%%%%%%%%%%%%%%%%%%%%%%%%%%%%%%%%%%%%%%%%%%%% 

\end{document}